%% file: ANA-HIGG-2017-10-PAPER.tex
\pdfoutput=1
\newcommand*{\ATLASLATEXPATH}{}
\documentclass[cernpreprint,  texlive=2016, UKenglish]{\ATLASLATEXPATH atlasdoc}
 
\usepackage[subfig]{\ATLASLATEXPATH atlaspackage}
\usepackage{\ATLASLATEXPATH atlasbiblatex}
\usepackage{adjustbox}
\usepackage[BSM]{\ATLASLATEXPATH atlasphysics}
 
\graphicspath{{logos/}{figures/}}
 
\usepackage{ANA-HIGG-2017-10-PAPER-defs}

\AtlasTitle{Search for scalar resonances decaying into \mumu in events with and without \btagged jets produced in proton--proton collisions at $\rts=13$ $\TeV$ with the ATLAS detector}

\AtlasAbstract{
A search for a narrow scalar resonance
decaying into an opposite-sign muon pair produced in events with and without \btagged jets is presented
in this paper. The search uses 36.1~$\ifb$ of \rts= 13 $\TeV$ proton--proton collision data recorded by the ATLAS experiment at the LHC.
No significant excess of events above the expected Standard Model background is observed
in the investigated mass range of 0.2 to 1.0~\TeV.
The observed upper limits at 95\% confidence level on the cross section times branching
ratio for \bquark associated production and gluon--gluon fusion are between 1.9 and 41 fb and 1.6 and 44 fb respectively, which is consistent with expectations.
}

\AtlasRefCode{HIGG-2017-10}
 
\PreprintIdNumber{CERN-EP-2018-300}

\AtlasJournalRef{JHEP 07 (2019) 117}
\AtlasDOI{10.1007/JHEP07(2019)117}

\AtlasCoverSupportingNote{Search for high mass dimuon signatures in b-tagged and b-veto events}{https://cds.cern.ch/record/2270631}

\AtlasCoverAnalysisTeam{Sinead Farrington, Kostas Nikolopoulos, Christopher McNicol, Elisabetta Pianori, Paul Thompson, Ligang Xia}
\AtlasCoverEdBoardMember{Jason Nielsen(chair), Vadim Kostyukhin, Paolo Mastrandrea}
 
\AtlasCoverEgroupEditors{atlas-HIGG-2017-10-editors@cern.ch}
 
\AtlasCoverEgroupEdBoard{atlas-HIGG-2017-10-editorial-board@cern.ch}
 
\hypersetup{pdftitle={ATLAS document},pdfauthor={The ATLAS Collaboration}}
 
\begin{document}
 
\maketitle

% The next lines are included from the .//ANA-HIGG-2017-10-PAPER-intro.tex input file
\section{Introduction}
\label{sec:intro}

This paper presents a search, in proton--proton (\pp) collision data at a centre-of-mass energy of 13 $\TeV$, targeting
a scalar particle $\X$ with a mass in the range 0.2--1.0~$\TeV$ decaying into two opposite-sign muons.
The natural width of the particle is assumed to be much narrower than the experimental resolution,
such that it is the latter that dominates the distribution of the dimuon invariant mass in signal events.
To maximize the sensitivity to the presence of \bquarks, the analysis is performed
in two data categories. The first search category requires at least one jet tagged as containing $b$-hadrons (\btagged) in the final state. The second category requires exactly zero $b$-tagged jets in the event.
Simultaneous fits to these two categories are used to search for both the gluon--gluon fusion and $b$-associated production mechanisms of the scalar particle.
This work is primarily motivated as a signature-based search; the event selection is not designed to target any specific model and instead facilitates the comparison of the data with predictions from various theoretical models.

The discovery of the 125 $\GeV$ Higgs boson at the Large Hadron Collider (LHC)
\cite{HIGG-2012-27,CMS-HIG-12-028} was a significant step in
understanding the mechanism of electroweak symmetry breaking (EWSB).
Measurements of the properties of this particle
\cite{HIGG-2014-06,HIGG-2013-01,CMS-HIG-14-009,CMS-HIG-14-018,HIGG-2015-07}
are so far consistent with expectations for the Standard Model (SM) Higgs boson
\cite{ENGLERT,HIGGS,HIGGS2,HIGGS3,Guralnik:1964eu,Kibble:1967sv}.
However, the observed Higgs boson could be only a part of an
extended scalar sector, as predicted by several theoretical models beyond the SM.
For example, such an extended scalar sector is predicted by a class of extensions of the SM known as the two-Higgs-doublet models (2HDM) \cite{Djouadi:2005gj,Branco:2011iw}.
 
\begin{figure}[!ht]
\centering
\vspace{5mm}
\subfloat[]{\includegraphics[height=0.2\textheight]{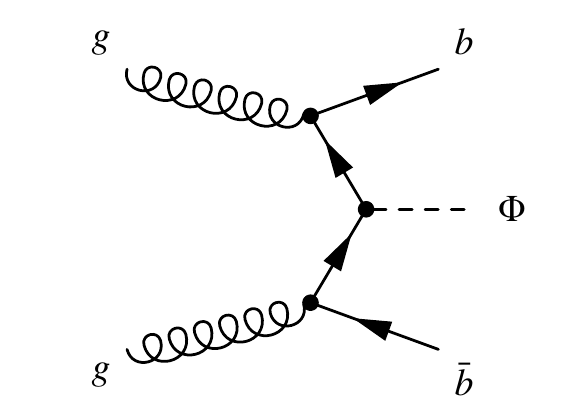}\label{fig:bbH_Feyn}}
\hspace{15ex}
\subfloat[]{\includegraphics[height=0.2\textheight]{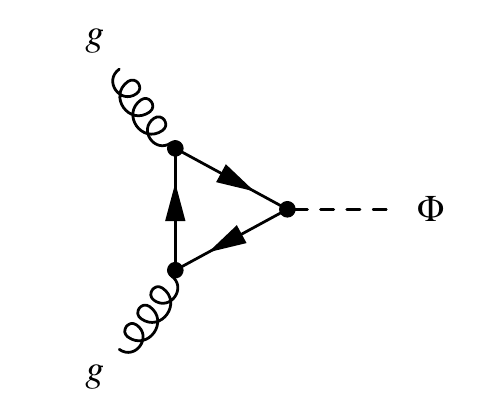}\label{fig:ggH_Feyn}}
\caption{Feynman diagrams for (a) the \bquark associated production mode and (b) the gluon-gluon fusion production mode.}
\label{fig:all_Feyn}
\end{figure}

The 2HDM posit the existence of three neutral bosons with properties and coupling strengths differing from those of the SM Higgs particle.
One of the higher-mass neutral Higgs bosons may couple to muons at a higher rate than to $\tau$-leptons, in contradiction to the SM Yukawa ordering. This is the case in the Flavourful Higgs model \cite{flavourfulhiggs}, for example.
The search described in this paper
sets limits on the cross-section times branching ratio of \X decaying into muon pairs which are more stringent
than the equivalent limits for \X decaying into $\tau$-lepton pairs \cite{HIGG-2016-12}.
This is due to higher identification efficiency and lower background rate
for muons than for $\tau$-leptons, and the invariant-mass resolution being better for muon pairs than for $\tau$-lepton pairs.
Similarly, the coupling of these higher-mass neutral Higgs bosons to \bquarks could be enhanced relative to the SM Higgs boson coupling. 
Hence, the production of \X in association with \bquarks (\bbx), shown in Figure~\ref{fig:bbH_Feyn}, could be more abundant than the production of \X by gluon--gluon fusion (ggF), shown in Figure~\ref{fig:ggH_Feyn}. The \Xbb decay mode, where \X is produced in associations with \bquarks, is investigated by CMS in Ref.~\cite{CMS-HIG-16-018}.
 
Additional interest for the search for an excess of events with respect to the SM predictions in the dimuon plus \btagged jets final state arises from $\Zprime$ and leptoquark models \cite{leptoquarkzprime}, which could result in either resonant or non-resonant contributions to the $pp \to t\overline{t}\mu^+\mu^- \to WbW\bar{b}\mu^{+}\mu^{-}$ final state. This work complements the search for $\Zprime$ candidates decaying into muon pairs presented in Ref.~\cite{EXOT-2016-05}:
by classifying events in terms of the presence or absence of a \btagged jet, it may be sensitive to signatures not observable with an inclusive selection.
 
Previous similar searches focused explicitly on beyond-the-SM Higgs bosons decaying into muon pairs, in the mass range 90--500~$\GeV$ \cite{HIGG-2012-11,CMS-HIG-13-024} using \rts= 7 and 8 $\TeV$ data. This work extends the exclusion limits for new scalar resonances with masses up to 1.0~$\TeV$.

The analysis described in this paper makes use of a fit to the observed dimuon invariant mass ($\mll$) spectra.
The presence of a signal produced either in association with \bquarks or via gluon--gluon fusion is tested with simultaneous fits to the two data regions (with and without a \btagged jet) separately for \bbx and ggF production modes. No assumptions are made about the relative contributions of the \bbx and ggF cross-sections.

This paper is structured as follows. Section \ref{sec:detector} describes the ATLAS detector. Section \ref{sec:samples} discusses the data and the simulated event samples used to model the signal and the background processes.
The event reconstruction is discussed in Section \ref{sec:analysis}, while Section \ref{sec:background} describes the background estimate and introduces the signal interpolation procedure used to model the $\mll$ distribution for resonance mass hypotheses for which no simulated sample was generated. The event yields and the description of the statistical analysis are discussed in Section \ref{sec:stat},
followed by Section \ref{sec:systematics}, which provides an overview of the systematic uncertainties.
Section \ref{sec:result} summarizes the results.

% End of text imported from the .//ANA-HIGG-2017-10-PAPER-intro.tex input file
% The next lines are included from the .//ANA-HIGG-2017-10-PAPER-detector.tex input file
\section{ATLAS detector}
\label{sec:detector}
 
ATLAS~\cite{PERF-2007-01},~\cite{ATLAS-TDR-19} is a multipurpose particle
detector with a forward--backward symmetric cylindrical
geometry and near 4$\pi$ coverage in solid angle.\footnote{ATLAS uses a
right-handed coordinate system with its origin at the nominal
interaction point (IP) in the centre of the detector and the
$z$-axis along the beam pipe. The $x$-axis points from the IP to the
centre of the LHC ring, and the $y$-axis points upward. Cylindrical
coordinates $(r,\phi)$ are used in the transverse plane, $\phi$
being the azimuthal angle around the $z$-axis. The pseudorapidity is
defined in terms of the polar angle $\theta$ as
$\eta=-\ln\tan(\theta/2)$. The angular distance is measured in units of $\Delta R=\sqrt{(\Delta\eta)^2+(\Delta\phi)^2}$.} It consists of an inner tracking detector surrounded by a thin superconducting
solenoid, electromagnetic and hadronic calorimeters, and a muon spectrometer.

The inner tracking detector (ID) covers the pseudorapidity range
$|\eta|<2.5$, and is surrounded by a superconducting solenoid
providing a \SI{2}{T} magnetic field.
At small radii, a high-granularity silicon pixel detector covers the
vertex region and typically provides four measurements per track.
It is followed by the silicon microstrip tracker, which provides eight
measurement points per track. The silicon detectors are complemented
by a gas-filled straw-tube transition radiation tracker,
which extends the tracking capability radially with typically 35 measurements per track for particles at $|\eta|=2.0$.
 
Electromagnetic (EM) calorimetry, within the region $|\eta|<3.2$, is
provided by barrel and endcap high-granularity lead/liquid-argon (LAr)
sampling calorimeters, with an additional thin LAr presampler covering
$|\eta| < 1.8$ to correct for energy loss in the upstream material.
For $|\eta| < 2.5$ the EM calorimeter is divided into
three layers in depth.
A steel/scintillator-tile calorimeter provides hadronic calorimetry for
$|\eta|<1.7$. LAr ionization, with copper as absorber, is
used for the hadronic calorimeters in the endcap region
$1.5<|\eta|<3.2$.  The solid-angle coverage is completed with forward
copper/LAr and tungsten/LAr calorimeter modules in $3.1 < |\eta| <
4.9$, optimized for EM and hadronic measurements,
respectively.
 
The muon spectrometer (MS) surrounds the calorimeters and comprises
separate trigger and high-precision tracking chambers measuring the
deflection of muons in a magnetic field provided by one barrel and two
end-cap toroid magnets. The precision chamber system covers the
region $|\eta| < 2.7$ with three layers of monitored drift tubes,
complemented by cathode-strip chambers in the forward region. The muon
trigger system covers the range $|\eta| < 2.4$ with resistive-plate
chambers in the barrel and thin-gap chambers in the endcap regions.
 
A two-level trigger and data acquisition system is used to record
events for offline analysis~\cite{TRIG-2016-01}. The level-1 trigger
is implemented in hardware and uses a subset of the detector information
to reduce the event rate to a value of at most \SI{100}{kHz}. It is followed
by a software-based high-level trigger which filters events using the
full detector information and outputs events for permanent storage at
an average rate of \SI{1}{kHz}.
 
% End of text imported from the .//ANA-HIGG-2017-10-PAPER-detector.tex input file
% The next lines are included from the .//ANA-HIGG-2017-10-PAPER-samples.tex input file
\section{Data and simulated event samples}
\label{sec:samples}
 
This search is performed with a sample of \pp collision data recorded
at a centre-of-mass energy \rts=13 $\TeV$ corresponding to an
integrated luminosity of 36.1~$\ifb$ collected during 2015 and 2016.
Events are considered for further analysis only if
they were collected under stable LHC beam conditions and the relevant
detector components were fully operational.
 
The data used in this analysis were collected using a combination of
muon triggers. For the 2015 dataset an isolated muon with transverse
momentum $\pt$ greater than $20\,\GeV$ was required, while for 2016
this requirement was raised to $26\,\GeV$. To avoid inefficiencies due
to the isolation requirement, these triggers were complemented by a
trigger requiring a reconstructed muon with $\pt$ greater than
$50\,\GeV$, without any isolation requirements.

Simulated signal events were generated similarly to the SM Higgs boson $H$, where $H$ was replaced by a higher-mass scalar boson $\X$ with a constant width of 4~\MeV. Nine Monte Carlo (MC) samples were generated in the mass range 0.2--1.0~$\TeV$ at intervals of 100~$\GeV$.  Both the ggF and \bbx production modes were considered.
Events were generated at next-to-leading order (NLO) with \POWHEGBOXV{2}~\cite{Nason:2004rx,Frixione:2007vw,Alioli:2010xd,Bag:2012} and
\progname{\MGMCatNLO}~\cite{Alwall:2014, Wiesemann:2014ioa},
for ggF and \bbx respectively. The CT10~\cite{Lai:2010vv} set of parton distribution
functions (PDFs) was used in the generation of ggF events
while \progname{CT10nlo\_nf4}~\cite{Dulat:2015mca} PDFs were used to produce
the \bquark associated signal samples in the four-flavour scheme.
In the gluon--gluon fusion production mode, \PYTHIA 8.210~\cite{Sjostrand:2014zea} with the AZNLO~\cite{STDM-2012-23}
set of tuned parameters was used together with the CTEQ6L1~\cite{Pumplin:2002vw}
PDF set for the parton shower, underlying event and hadronization simulation.
For the \bquark associated production, \PYTHIA 8.210 with the A14~\cite{ATL-PHYS-PUB-2014-021}
set of tuned parameters was used together with the NNPDF2.3LO~\cite{Ball:2012cx} PDF set
for the parton shower, underlying event, and hadronization simulation.
The EvtGen v1.2.0 program~\cite{Lange:2001uf}  was
used to model the properties of the bottom and charm hadron decays in all signal samples.

Since the generated signal samples are spaced equally in $\mx$ while the experimental resolution increases with $\mx$, signal distributions for intermediate mass hypotheses were obtained using an interpolation procedure, described in Section~\ref{sec:background}.

Events containing \Zjets were simulated using \POWHEGBOXV{2}~\cite{Alioli:2010xd, Alioli:2010qp} interfaced to the \PYTHIA~8.186
parton shower model and the CT10 PDF set. The AZNLO tune was used, with
PDF set CTEQ6L1 for the modelling of non-perturbative effects. The EvtGen v1.2.0 program was
used to model the properties of the bottom and charm hadron decays. Photos++ 3.52~\cite{Davidson:2010ew} was used for photons radiated from electroweak vertices and charged leptons.
Generator-level filters are employed to enhance the fraction of simulated events in the phase-space region that is most relevant for the analysis.
Event yields were corrected with
a mass-dependent rescaling at next-to-next-to-leading order (NNLO) in the QCD coupling constant, computed with \progname{VRAP} 0.9~\cite{Anastasiou:2003ds} and the
CT14NNLO PDF set~\cite{Dulat:2015mca}. Mass-dependent electro-weak (EW) corrections were computed at NLO with 
{\texttt{mcsanc}}--1.20~\cite{Bondarenko:2013nu},
along with photon-induced contributions ($\gamma\gamma\to \ell\ell$ via $t$- and $u$-channel processes) computed with
the MRST2004QED PDF set~\cite{Martin:2004dh}.
 
Additional \Zjets events, used to evaluate the systematic uncertainty from the modelling of the fitted \mll distribution,
were simulated with \progname{Sherpa}~2.2.1 ~\cite{SHERPA1.1} and with \progname{\MGMCatNLO} interfaced with \progname{Pythia 8.186}.
In \progname{Sherpa}, \Zjets matrix elements were evaluated with up to two extra partons at NLO,
and three or four extra partons were included at LO in QCD. The merging of different parton multiplicities was achieved through a matching scheme
based on the CKKW-L~\cite{Lonnblad:2001iq}~\cite{Lavesson:2005xu} merging technique using a merging scale of $Q_{\text{cut}} = 20$~\GeV.
The models used for the parton shower and underlying event are the ones provided internally by \progname{Sherpa}.
The \progname{Sherpa}~2.2.1 generator adopts a full five-flavour scheme, with massless \bquarks and \cquarks in the matrix elements. These quarks are given mass in the final state and massive quarks can be produced directly via the parton shower.
The \progname{\MGMCatNLO} v2 generator merges the LO (QCD) matrix-element calculations for \Zjets
with up to four additional partons (higher jet multiplicities are modelled by the parton shower algorithm). The merging scheme applied to
combine different parton multiplicities is the CKKW-L scheme with a merging scale of $Q_{\text{cut}} = 30$~\GeV.
For the LO matrix-element calculation the NNPDF2.3 LO PDFs were used (with \alphas = 0.13).
The A14 tune, together with the NNPDF2.3 LO PDFs, was used for the parton shower.
Similarly to \progname{Sherpa}~2.2.1, \progname{\MGMCatNLO} adopts a full five-flavour scheme.

For the generation of \ttbar and single top-quark production in the $Wt$- and $t$-channel, the \POWHEGBOXV{2} generator with the CT10 PDF set was used for the matrix element calculations.
The parton shower, fragmentation, and the underlying event were simulated using
\PYTHIA 6.428 with the CTEQ6L1 PDF set and the corresponding Perugia 2012 tune~\cite{Skands:2010ak}. The top-quark mass
was set to $172.5\,\GeV$.
For the generation of \ttbar events, the \hdamp parameter of \POWHEGBOX, which controls the transverse momentum of the first additional emission beyond the Born configuration, was set to the mass of the top quark. The main effect of this parameter is to regulate the high transverse momentum emission against which the \ttbar system recoils.
The $t\bar{t}$ production sample was normalized to the predicted production cross-section as
calculated with the \progname{Top++2.0} program to NNLO in perturbative QCD, including soft-gluon resummation
to next-to-next-to-leading-log (NNLL) order~\cite{Czakon:2011xx}. The normalization of the single
top-quark event samples used an approximate calculation at NLO in QCD for the $t$-channels~\cite{Aliev:2010zk, Kant:2014oha} while NLO+NNLL predictions were used for the $Wt$-channel~\cite{Kant:2014oha}. The EvtGen v1.2.0 program was
used to describe the properties of the bottom and charm hadron decays.

Diboson processes were modelled using the \progname{Sherpa}~2.1.1 ~\cite{SHERPA1.1}
generator and they were calculated for up to one ($ZZ$) or zero ($WW$, $WZ$)
additional partons at NLO and up to three additional partons at leading order (LO) using
the Comix~\cite{Gleisberg:2008fv} and OpenLoops~\cite{Cascioli:2011va} matrix-element generators. The matrix element calculation was merged with the
\progname{Sherpa} internal parton shower~\cite{Schumann:2007mg} using the ME+PS@NLO prescription~\cite{Hoeche:2012yf}. The CT10 PDF
set was used in conjunction with dedicated parton shower tuning
developed by the \progname{Sherpa} authors. The generator cross-sections, calculated at NLO, were used
in this case.

The generated signal samples were simulated with the
fast simulation~\cite{SOFT-2010-01,ATL-PHYS-PUB-2010-013}
framework of ATLAS, which replaces the full simulation of the electromagnetic and hadronic calorimeters by a parameterized model. The background samples were simulated using
the full Geant4-based simulation of the
detector~\cite{Agostinelli:2002hh}. Finally, the simulated events were
processed through the same reconstruction software as the data.  The
effects of pile-up from multiple proton--proton interactions in the same and
neighbouring bunch crossings were accounted for by overlaying minimum-bias events simulated using \PYTHIA8 with the A2 tune~\cite{ATL-PHYS-PUB-2012-003} and interfaced
with the MSTW2008LO PDFs. Simulated events were then reweighted to match the
pile-up conditions observed in the data.
% End of text imported from the .//ANA-HIGG-2017-10-PAPER-samples.tex input file
% The next lines are included from the .//ANA-HIGG-2017-10-PAPER-analysis.tex input file
\section{Event reconstruction}
\label{sec:analysis}
 
Interaction vertices are reconstructed~\cite{PERF-2015-01} from
tracks measured by the inner detector. The primary vertex is defined as the vertex with the
largest $\sum \pT^2$ of its associated tracks. It must have at least two associated tracks,
each with transverse momentum $\pT > 400\,\MeV$.
 
Muon candidates are reconstructed from an ID track combined with
a track or track segment detected in the muon spectrometer~\cite{PERF-2015-10}. These two tracks are used as inputs
to a combined fit (for $\pt$ less than 300 \GeV) or to a statistical
combination (for $\pt$ greater than 300 \GeV)~\cite{PERF-2015-10}.
The combined fit takes
into account the energy loss in the calorimeter and
multiple-scattering effects. The statistical combination 
for high transverse momenta is performed to mitigate the effects of relative ID
and MS misalignments.
High-$\pT$ muon candidates are required to have at least three hits, one in each of the three layers of precision chambers in the MS. For these muons, specific regions of the MS where the alignment is suboptimal are vetoed as a precaution, as well as the transition region between the MS
barrel and endcap ($1.01 < |\eta| < 1.10$).
Further quality
requirements are applied to the consistency of the ID and MS momentum
measurements, and to the measured momentum
uncertainty for the MS track.

The reconstructed muon candidates are required to have transverse
momentum, $\pt$, greater than $30\,\GeV$ and pseudorapidity
$|\eta|<2.5$. They are further required to be consistent with the hypothesis that they originate
from the primary vertex by applying selections on the transverse
($d_{0}$) and longitudinal ($z_{0}$) impact parameters, defined
relative to the primary vertex position: $|d_{0} / \sigma_{d_{0}} |<3$
and
$| z_{0} \sin{\theta} | < 0.5$ mm
where $\sigma_{d_{0}}$ denotes the uncertainty in the transverse impact
parameter. A loose isolation requirement is applied, based on the
sum of the momenta of inner detector tracks which lie within a
variable-sized cone, with $\Delta R_{\max} = 0.3$, around the muon
track. This isolation requirement is tuned to yield a 99\% efficiency over
the full range of muon $\pt$.

Jets are reconstructed from noise-suppressed energy clusters in the
calorimeter~\cite{PERF-2014-07}  with the \antikt
algorithm~\cite{Cacciari:2008gp, Cacciari:2011ma} with radius
parameter $R = 0.4$.
The energies of the jets are calibrated using a
jet energy scale (JES) correction derived from both simulation and in
situ studies using data~\cite{
PERF-2016-04}.
All jets are required to have $\pt > 25\,\GeV$ and $|\eta| < 4.5$.
A multivariate selection that reduces contamination from pile-up~\cite{PERF-2014-03} is applied to jets with $\pt < 60$ \GeV~and with $|\eta| < 2.4$ utilizing calorimeter and tracking information to separate hard-scatter jets from pile-up jets.

Selected jets in the central region can be tagged as containing
$b$-hadrons (\btagged) by using a multivariate discriminant
(MV2c10)~\cite{PERF-2012-04, ATL-PHYS-PUB-2016-012} that combines
information from an impact-parameter-based algorithm, from the
explicit reconstruction of a secondary vertex and from a multi-vertex
fitter that attempts to reconstruct the full $b$- to $c$-hadron decay
chain. At the chosen working point, the algorithm provides nominal
light-flavour ($u$,$d$,$s$-quark and gluon) and \cjet misidentification
rates of 3\% and 32\%, respectively, for an average 85\%
\bjet tagging efficiency, as estimated from simulated $t\bar{t}$
events for jets with $\pt > 20\,\GeV$ and $|\eta| < 2.5$.
The flavour-tagging efficiencies from simulation are corrected
separately for $b$-, $c$- and light-flavour jets, based on the respective
data-based calibration analyses~\cite{PERF-2016-05}.
 
Simulated events were corrected to reflect the muon and jet momentum scales and resolutions,
\bjet identification algorithm calibration, as well as the muon trigger, identification, and isolation efficiencies measured in data.

An overlap removal procedure is applied to avoid a single particle being reconstructed as two different objects
as follows.
Jets not tagged as $\bjets$, but which are reconstructed within $\Delta R = 0.2$ of a muon, are removed if they have fewer than three associated tracks or if the muon energy constitutes more than 50$\%$ of the jet energy. Muons reconstructed within a cone of
size $\Delta R = \min\left(0.4, 0.04 + 10\,\GeV/\pt\right)$ around the jet axis of any surviving jet are removed.
Jets are also discarded if they are within a cone of size $\Delta R=0.2$ around an electron candidate.
Electrons are reconstructed from clusters of energy deposits in the electromagnetic calorimeter that match ID tracks, and are identified as described in Ref.~\cite{ATLAS-CONF-2016-024}. For electrons with transverse energy $\et > 10\,\GeV$ and pseudorapidity of $|\eta|$ < 2.47, a likelihood-based selection is used at the “loose” operating point defined in~\cite{ATLAS-CONF-2016-024}.
 
Following the overlap removal, jet cleaning criteria are
applied to identify jets arising from non-collision sources or noise in
the calorimeters, and any event containing such a jet is
removed~\cite{ATLAS-CONF-2015-029}.
 
The missing transverse momentum, $E_{\text{T}}^{\text{miss}}$, is defined as the negative vector sum of the transverse
momenta of muons, electrons and jets associated with the primary vertex.
A soft term~\cite{PERF-2016-07} is added
to include well-reconstructed tracks matched to the primary vertex that are not associated with any of the objects.
 
At least two reconstructed muons are required in the event, and the
two highest-$\pt$ muons of the event are used to form a dimuon
candidate. If these muons do not have opposite charges, the event is
rejected.
 
% End of text imported from the .//ANA-HIGG-2017-10-PAPER-analysis.tex input file
% The next lines are included from the .//ANA-HIGG-2017-10-PAPER-bkg.tex input file
\section{Signal and background estimate}
\label{sec:background}

Events satisfying the preselection criteria of Section \ref{sec:analysis} are considered for further analysis.
Depending on the value of the reconstructed dimuon invariant mass and the number of \btagged jets, they are classified as being in either a control region, used to measure the rate of the dominant backgrounds
(\Zjets and \ttbar) or a signal region, used to search for the $\X\to\mu\mu$ signal.
All signal and control regions are designed to be orthogonal.
 
Selected events are retained in the signal region if the dimuon
invariant mass, $\mll$, exceeds $160$~\GeV, while they are retained in the control region if
$100\,\GeV<\mll<160\,\GeV$. The control regions do not include the $Z$-boson peak; this
avoids constraints on systematic uncertainties in a region of phase space
which may behave differently with respect to the high mass tails where the signal search is focused.
 
Events in the signal region are further
classified as ``SRbTag'', if at least one \btagged jet is identified in the event, or ``SRbVeto'', otherwise.
No further optimization of the signal region definition is considered, in order to minimize the assumptions made about the signal kinematics.
 
Events in the control region are further classified as ``CRbVeto'' if there is no \btagged jet
identified in the event, while remaining control region events are classified as ``CRbTag'', if the missing transverse momentum is less
than $100\,\GeV$, or ``CRttbar'' otherwise.
 
The dominant background sources for this signature are muon pairs through  \Zjets, top-quark pair, and diboson production.
$W$+jets and QCD multi-jet events contribute less than 0.01$\%$ to the total background in signal regions, and therefore are neglected.
All relevant background contributions were modelled
using simulated samples 
as described in Section~\ref{sec:samples}.
 
Simulated \Zjets events were categorized depending on the generator-level ``truth'' labels of the
jets in the event. Simulated jets were truth-labelled according to which hadrons with $\pt >5\,\GeV$ were found within a cone of size $\Delta R = 0.3$ around the
jet axis. If a $b$-hadron was found, the jet was truth-labelled as a \bjet. If no
$b$-hadron was found, but a $c$-hadron was present, then the jet was truth-labelled
as a \cjet. Otherwise the jet was truth-labelled as a light-flavour (i.e.,
$u$,$d$,$s$-quark, or gluon) jet.
The \Zjets events were classified as $\Zboson\text{+\,heavy flavour}$ ($\ZHF$)
if a \bjet or \cjet was found at generator level and $\Zboson\text{+\,light flavour}$ ($\ZLF$) otherwise.
These two categories of simulated events were treated as different background components.
 
The predictions for the expected numbers of $\ttbar$, $\ZHF$ and $\ZLF$ events are adjusted to match the number of events in data in the control regions via a global fit (see Section \ref{sec:stat}), while the other backgrounds are normalized to their expected cross-section, discussed in Section~\ref{sec:samples}.
 
The  expected  invariant  mass  distribution  of  the
narrow resonance signal $\mx$ is modelled with a double-sided Crystal Ball (DSCB)~\cite{HIGG-2016-08}
function in the mass range 0.2 $\TeV <\mx<$ 1.0 $\TeV$ for all nine simulated MC samples.
The width of the $\mll$ distribution is dominated by the experimental resolution, which can be described by a Gaussian distribution.
The power-law asymmetric terms of the DSCB have enough degrees of freedom to model the \mll tails for signals in the 0.2--1.0 TeV mass range. In order to interpolate the signal parameterization to any mass value in this fit range,
second-order polynomial parameterizations of all six signal-shape parameters as a function of $\mx$ are obtained from a simultaneous
fit to all the generated mass points $\mx$, separately for events with and without at least one \btagged jet.
Since all signal samples are scaled to the same arbitrary cross-section, 
any differences in the number of events are due to differences in the acceptances of the selection.
The fitted normalization is parameterized with a second-order polynomial in a similar way to the other DSCB parameters.
As a result of this procedure, the \mll distribution for an arbitrary hypothesis mass can be found by constructing a DSCB from parameters calculated using the fitted coefficients and the hypothesized mass \mx. The interpolation is performed separately for each systematic variation, and for the gluon--gluon fusion and \bquark associated production signal samples. Binned templates are then generated from the DSCB.
To validate the interpolation procedure, one simulated sample at a time is removed from the simultaneous fit, and the template generated from the DCSB is compared with the \mll distribution from the corresponding simulated sample and they agree within 2--5$\%$, the size of the bin-by-bin MC statistical uncertainty.
Nine MC samples were generated in the mass range 0.2--1.0 \TeV, every 100 \GeV. The interpolation procedure was used to generate signal templates every 10 $\GeV$ between 0.2 and 0.3 $\TeV$, every 20 $\GeV$ between 0.3 and 0.6 $\TeV$, and every 50 $\GeV$ between 0.6 and 1.0 \TeV.
 
The acceptance of \bquark associated production in the SRbTag region varies in the range 11--19$\%$ depending on \mx. The acceptance in the SRbVeto region, which is inversely correlated with the SRbTag one, varies from 20$\%$ to 15$\%$. For this reason, both SRbTag and SRbVeto are included in the global fit (see Section~\ref{sec:stat}). The acceptance of gluon--gluon fusion in the SRbVeto region varies in the range 31--35$\%$. The acceptance in the SRbTag is less than 2$\%$ for all masses.

% End of text imported from the .//ANA-HIGG-2017-10-PAPER-bkg.tex input file
% The next lines are included from the .//ANA-HIGG-2017-10-PAPER-fit.tex input file
\section{Statistical analysis}
\label{sec:stat}

The test statistic $\tilde{q}_\mu$ as defined in Ref.~\cite{Cowan:2010js} is used to determine
the probability that the background-only model is compatible with the observed data,
to extract the local $p$-value, and, if no hint of a signal is found in this procedure, to derive exclusion intervals using
the CL$_\text{s}$ method.
The binned likelihood function is built as the product of Poisson probability terms associated with the bins in the \mll distribution in the 0.16--1.5~$\TeV$ range.
It depends on the parameter of interest, on the normalization factors of the dominant backgrounds and on additional nuisance parameters (representing the estimates of the systematic uncertainties) that are each constrained by a Gaussian prior. The bin-by-bin statistical uncertainties of the simulated backgrounds are also considered.
Separate fits are performed to test the different hypotheses: only \bbx signal, only ggF signal, and signal composed of different mixtures of the two production modes. Data are fitted in the 0.16--1.5~$\TeV$ range to check that the background description is correct and to constrain systematic uncertainties, while the signal presence is tested in a narrower 0.2--1.0~$\TeV$ range.
 
Logarithmic binning in $\mll$ is chosen to scale with experimental dimuon mass resolution (defined as full-width at half-maximum), which varies from 5\% at $\mll=200\,\GeV$ to 14\% at $\mll=1\,\TeV$, while ensuring that the number of simulated background events in each bin is sufficient to reliably predict the background.
The numbers of events in the CRbVeto, CRbTag, and CRttbar control regions are also included in
the maximum-likelihood fit. Data in these regions are used to constrain the
normalizations of the dominant background processes: $\ZLF$,
$\ZHF$, and $t\bar{t}$. The normalization factors used to scale
the expected number of events for each process to match the data are free parameters in the fit.
Their expected uncertainty is estimated using an Asimov dataset \cite{Cowan:2010js}, a pseudo-data distribution equal
to the background plus signal expectation for a given value of the signal cross-section times branching ratio.
The expected and measured normalization factors are shown in Table~\ref{tab:normFitResults}.
They are measured in a fit to data under the background plus signal hypothesis, where the signal corresponds to \bbx production with mass \mx= 480 \GeV.
 
\begin{table}[!ht]
\caption{
Normalization factors of the dominant backgrounds as measured in a fit to data under
the background plus signal hypothesis (480 GeV \bbx signal). The uncertainty includes both the statistical and systematic
sources: the latter ones dominate. The reduction of the observed relative uncertainty of the $\ZHF$ normalization in the data fit is due to
the large increase in the number of $\ZHF$ events
when the data in signal regions are included.}
\begin{center}
\label{tab:normFitResults}
\begin{tabular}{  c | c | c}
\hline\hline
Background &  Asimov data & Observed data\\ \hline 
$\ZHF$    & $1.00 \pm 0.23$ &$ 1.47 \pm  0.26  $ \\ \hline
$\ZLF$    & $1.00\pm 0.02$ &$ 1.02 \pm  0.02 $ \\ \hline
$t\bar{t}$ & $1.00\pm0.04$ &$ 1.02 \pm  0.04  $\\
\hline \hline
\end{tabular}
\end{center}
\end{table}

As discussed in Section~\ref{sec:systematics}, the experimental uncertainties are treated
as fully correlated among different processes and regions, while the
theoretical uncertainties are mostly uncorrelated among processes.
 
The numbers of observed events in the signal regions together with the
predicted event yields from signal and background processes are shown
in Table ~\ref{tab:yields_postfit}. The numbers are the results of the maximum-likelihood fit to data under the background plus signal hypothesis, where the signal corresponds to \bbx production with \mx= 480 \GeV.
The number of predicted signal events corresponds to the expected upper limit (UL) times the cross-section:
this highlights the comparison between signal and background at the edge of the analysis sensitivity.
 
The observed numbers of events are compatible with those expected from SM processes, within uncertainties.
The $\mll$ distribution in the SRbTag region is shown in Figure~\ref{Cs_SRbTag} for a \bbx-only fit.
The $\mll$ distribution in the SRbVeto region is shown in Figure~\ref{Cs_SRbVeto}, for a ggF-only fit.
Each background process is normalized according to its post-fit cross-section. The templates for the \mx=200 $\GeV$, \mx=480 $\GeV$ and \mx=1 $\TeV$
mass hypotheses are normalized to the expected upper limit.

\begin{table}[!ht]
\caption{Post-fit numbers of events from the combined fit under the background plus signal hypothesis ($480\,\GeV$ \bbx signal). Here ``Total Bkg'' represents the sum of all backgrounds. The quoted uncertainties are the combination of statistical and systematic uncertainties. The uncertainty in the total background determined by the fit is smaller than the sum in quadrature of the individual components due to the normalization factors and systematic uncertainties that introduce correlation between them. The number of signal events corresponds to the expected upper limit (UL) times the cross-section. The uncertainties in the expected signal yields are from fits using the Asimov dataset. \label{tab:yields_postfit}}
\renewcommand{\arraystretch}{1.3}
\begin{center}
\adjustbox{max width=1\textwidth}{
\newcolumntype{C}{ @{}>{${}}c<{{}$}@{} }
\begin{tabular}{ l rCl rCl rCl rCl rCl }
\hline\hline
Sample & \multicolumn{3}{c}{SRbTag} & \multicolumn{3}{c}{SRbVeto} & \multicolumn{3}{c}{CRbTag} & \multicolumn{3}{c}{CRbVeto} & \multicolumn{3}{c}{CRttbar}\\
\hline
$t\bar{t}$ & 16\,490& \pm& 240 & 2\,090  & \pm& 300 & 16\,300 & \pm & 600  & 2\,320    & \pm & 350  & 4\,160 & \pm & 70 \\
Single top & 1\,470  & \pm& 100 & 480   & \pm& 50  & 1\,340   & \pm & 100  & 530     & \pm & 50   & 297  & \pm & 22 \\
Diboson   & 176   & \pm& 23  & 2\,570  & \pm& 180 & 280    & \pm & 40   & 4\,550    & \pm & 310  & 19   & \pm & 5 \\
$\ZHF$     & 1\,920  & \pm& 320 & 3\,400  & \pm& 700 & 13\,300 & \pm & 1\,900 & 25\,000  & \pm & 5\,000 & 18   & \pm & 9 \\
$\ZLF$     & 1\,060  & \pm& 330 & 57\,700& \pm& 900 & 6\,300   & \pm & 2\,000 & 501\,000 & \pm & 5\,000 & 10   & \pm & 9 \\
Total Bkg  & 21\,110& \pm& 140 & 66\,240& \pm& 250 & 37\,530 & \pm & 200  & 533\,100 & \pm & 700  & 4\,500 & \pm & 60 \\
\hline
\bbx (UL)& \multicolumn{3}{c}{$37_{-10}^{+14}$} & \multicolumn{3}{c}{$45_{-13}^{+18}$} & \multicolumn{3}{c}{}&\multicolumn{3}{c}{} &\multicolumn{3}{c}{} \\  
ggF (UL) & \multicolumn{3}{c}{$4_{-1}^{+2}$} & \multicolumn{3}{c}{$73_{-21}^{+28}$} & \multicolumn{3}{c}{} & \multicolumn{3}{c}{}& \multicolumn{3}{c}{}\\ 
\hline
Data       & \multicolumn{3}{c}{21\,154}  & \multicolumn{3}{c}{66\,300}  & \multicolumn{3}{c}{37\,527}  & \multicolumn{3}{c}{533\,134}  & \multicolumn{3}{c}{4\,511}  \\

\hline\hline
\end{tabular}
}
\end{center}
\end{table}

\begin{figure}[!ht]
\centering
\subfloat[]{\includegraphics[width=0.49\linewidth]{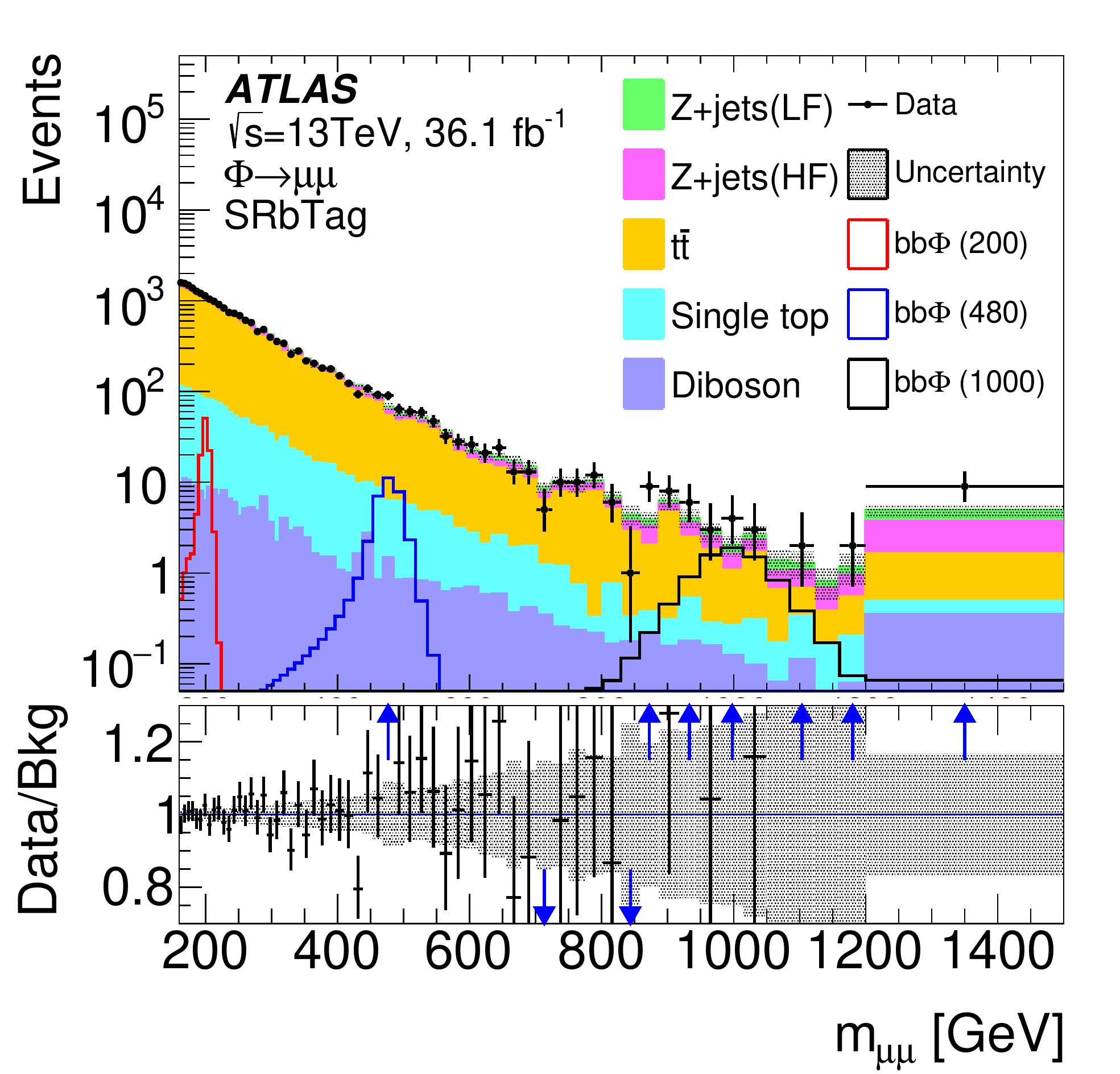} \label{Cs_SRbTag} }
\subfloat[]{\includegraphics[width=0.49\linewidth]{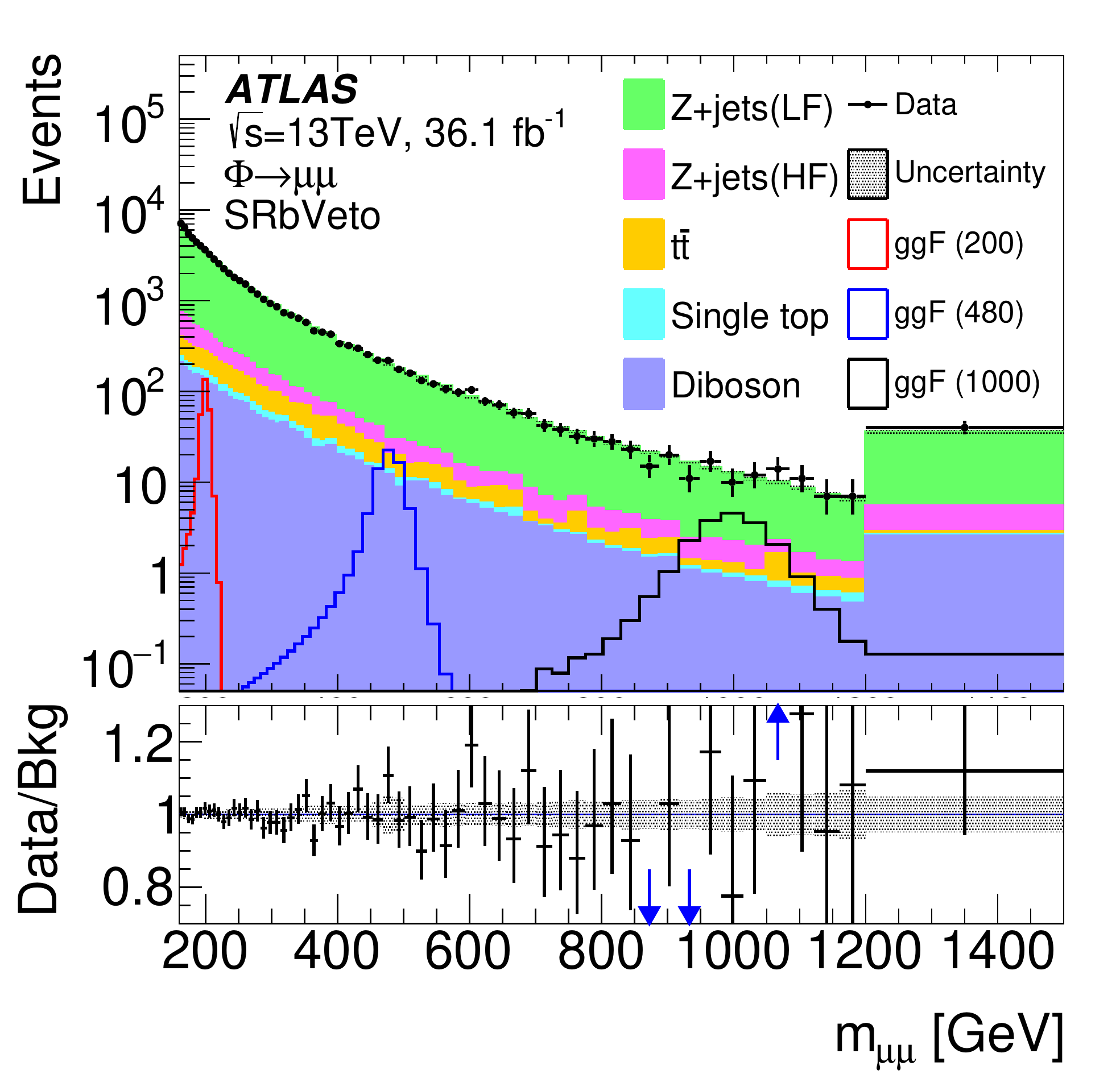}\label{Cs_SRbVeto}}
\caption{Distributions of the dimuon invariant mass, \mll, after the combined fit to data are shown normalized: (a) in the SRbTag, under the background plus signal hypothesis ($480\,\GeV$ \bbx signal) and (b) in the SRbVeto, under the background plus signal hypothesis ($480\,\GeV$ ggF signal). The fit for a \mx=480 $\GeV$ signal corresponds to the largest excess observed above the background expectation.
Each background process is normalized according to its post-fit cross-section. The templates for the \mx=200 $\GeV$, \mx=480 $\GeV$ and \mx=1 $\TeV$ mass hypotheses are normalized to the expected upper limit. The data are shown by the points,
while the size of the statistical uncertainty is shown by the error bars. The last bin includes the overflow.
The blue arrows represent the data points outside of the frame.
The hatched band shows the total systematic uncertainty of the post-fit yield.
}
\label{fig:postfitmllSR}
\end{figure}
 
% End of text imported from the .//ANA-HIGG-2017-10-PAPER-fit.tex input file
% The next lines are included from the .//ANA-HIGG-2017-10-PAPER-systematics.tex input file
\section{Systematic uncertainties}
\label{sec:systematics}
 
The sources of systematic uncertainty can be divided into three
groups: those of experimental origin, those related to the modelling
of the simulated backgrounds, and those associated with signal simulation.
The finite size of the simulated background samples
is also an important source of uncertainty.

\subsection{Experimental uncertainties}
\label{sec:expSyst}
 
The dominant experimental uncertainties originate from residual mismodelling of the muon reconstruction and selection after the simulation-to-data efficiency correction factors have been applied.
They include the uncertainty obtained from $Z\to\mu\mu$ data studies and a high-$\pt$ extrapolation uncertainty corresponding to the decrease in the muon reconstruction and selection efficiency with increasing $\pt$ which is predicted by the MC simulation. The degradation of the muon reconstruction efficiency was found to be approximately $3\%$ per~$\TeV$  as a function of muon $\pt$.
Uncertainties in the isolation and trigger efficiencies of muons~\cite{PERF-2015-10}, along with the uncertainty in their energy scale and resolution, are estimated from $Z\to\mu\mu$ data taken at \rts=13 $\TeV$. These are found to have only a small impact (they account for $<0.4\%$ of the total uncertainty on the fitted value of the signal cross-sections). Other sources of experimental uncertainties are the residual differences in the modelling of the flavour-tagging algorithm, the jet energy scale and the jet energy resolution after the
simulation-to-data correction factors are applied.
Flavour-tagging simulation-to-data efficiency correction
factors are derived~\cite{PERF-2012-04} separately for \bjets, \cjets,
and light-flavour jets. All three correction factors depend on jet
$\pt$ (or $\pt$ and $|\eta|$) and are affected by uncertainties from multiple sources.
These are decomposed into uncorrelated components
which are then treated independently, resulting in three uncertainties
for \bjets and for \cjets, and five for light-flavour jets. The
approximate size of the uncertainty in the tagging efficiency is 2\%
for \bjets, 10\% for \cjets and 30\% for light-flavour jets. Additional
uncertainties are considered in the extrapolation of the \bjet efficiency calibration above $\pt = 300\,\GeV$ and in the
misidentification of hadronically decaying $\tau$-leptons as
\bjets. The uncertainties in the jet energy scale and resolution are
based on their respective measurements in
data~\cite{PERF-2016-04}.
The many sources of uncertainty
in the jet energy scale correction are decomposed into 21 uncorrelated
components which are treated as being independent of one another.
An additional uncertainty that specifically affects the energy calibration of $\ensuremath{b}$-
and \cjets is considered.
 
The uncertainties in the energy scale and resolution of the
jets and leptons are propagated to the calculation of $E^{\text{miss}}_{\text{T}}$, which
also has additional uncertainties from the scale, resolution, and
efficiency of the tracks used to define the soft term, along with
the modelling of the underlying event.
 
The pile-up modelling uncertainty is assessed by varying the number of pile-up interactions in simulated events. The variations are designed to cover the uncertainty in the ratio of the predicted and measured cross-section
of non-diffractive inelastic events producing a hadronic system of mass $\mHadronicSystem > 13$~\GeV~\cite{STDM-2015-05}.
 
The uncertainty in the combined 2015+2016 integrated luminosity is 2.1$\%$. It is derived, following a methodology similar to that detailed in Ref.~\cite{DAPR-2013-01}, and using the LUCID-2 detector for the baseline luminosity measurements~\cite{LUCID2}, from calibration of the luminosity scale using $x$--$y$ beam-separation scans.

\subsection{Theoretical uncertainties}
\label{sec:theoSyst}
 
\subsubsection*{Background processes}
 
As discussed earlier, the rates of major backgrounds are measured using data control regions, thus theoretical uncertainties in the predicted cross-section for \Zjets and $t\bar{t}$ processes are not considered. Instead, the effects of theoretical uncertainties in the modelling of the \mll distribution and in the ratio of the event yields in signal regions to those in the control regions are evaluated.
The dominant theoretical uncertainties are the shape uncertainty
in the dimuon mass spectra of the $t\bar{t}$ and $\ZHF$ contributions.

For both the $\ZLF$ and $\ZHF$ processes, the following sources of theoretical and modelling
uncertainties are considered:
PDF uncertainties (estimated via eigenvector variations and by comparing different PDF sets), limited accuracy of the fixed-order calculation
(estimated by QCD scale variations),
variations in the choice of strong coupling constant value ($\alphas(M_{Z})$), EW corrections,
and photon-induced corrections. These variations are treated as fully correlated between $\ZLF$ and $\ZHF$ processes.
 
The PDF variation uncertainty is obtained using the $90\%$ confidence level (CL) CT14NNLO
PDF error set and by following the procedure described in
Refs.~\cite{Gao:2013bia, Butterworth:2015oua}. Rather than a
single nuisance parameter to describe the 28 eigenvectors of this PDF
error set, which could lead to an underestimation of its effect, a
re-diagonalized set of 7 PDF eigenvectors was
used~\cite{Dulat:2015mca}. This represents the minimal set of PDF eigenvectors that
maintains the necessary correlations, and the sum in quadrature of
these eigenvectors matches the original CT14NNLO error envelope
well. They are treated as separate nuisance parameters.
The uncertainties due to the variation of PDF scales and \alphas are derived using VRAP. The former is obtained by
varying the renormalization and factorization scales of the nominal
CT14NNLO PDF up and down simultaneously by a factor of two. The value
of \alphas used (0.118) is varied by $\pm 0.002$.  The EW
correction uncertainty was assessed by comparing the nominal additive
($1+\delta_\mathrm{EW}+\delta_\mathrm{QCD}$) treatment with the multiplicative
approximation (($1+\delta_\mathrm{EW}$)($1+\delta_\mathrm{QCD}$)) treatment of the EW
correction in the combination of the higher-order EW and QCD
effects. The uncertainty in the photon-induced correction is
calculated from the uncertainties in the quark masses and the
photon PDF.
Following the recommendations of the PDF4LHC
forum~\cite{Butterworth:2015oua}, an additional uncertainty due to the
choice of nominal PDF set is derived by comparing the central values
of CT14NNLO with those from other PDF sets, namely
MMHT14~\cite{Motylinski:2014sya} and NNPDF3.0~\cite{Ball:2014uwa}.
The maximum width of the envelope of these comparisons is used as the PDF choice uncertainty,
but only if it is larger than the width of the
CT14NNLO PDF eigenvector variation envelope.
 
An additional modelling uncertainty is considered for the $\ZHF$ process.
The \mll spectrum simulated by \POWHEG was compared with those simulated with
\progname{Sherpa}~2.2.1 ~\cite{SHERPA1.1} and with \progname{\MGMCatNLO} interfaced with \progname{Pythia 8.186}.
A functional form was chosen to describe the envelope of the differences in the \mll distribution shape between events with at least one \bquark simulated by \progname{Sherpa} 2.2.1 and \progname{\MGMCatNLO} v2.
 
These systematic uncertainties not only
imply uncertainties in the shape of the \mll distribution in the signal region, but they also affect the ratio of the expected number of events in the signal region to the expected number in the
control region. The effect on the $\ZLF$ normalization in the bVeto signal region is ~$2\%$, while the size of the uncertainty for $\ZHF$ in the bTag signal region is $7\%$.
 
Moreover, the ratio of the expected number of $\ZLF$ events in the bVeto control region to the events in bTag control and signal regions was estimated with the nominal \POWPYTHIA $Z$+jet sample, and with the alternative \progname{\MGMCatNLO} v2 sample. To cover the observed difference between the two, an additional uncertainty of $27\%$ in the normalization of the $\ZLF$ process in the bTag signal and control region was applied.

For the $t\bar{t}$ process, theoretical uncertainties in the modelling of the \mll spectrum and in the extrapolation from control region to signal region are also considered. These are estimated by comparing the nominal prediction with alternative ones. To estimate the impact of initial and final state radiation modelling, two alternative \POWPYTHIA samples were generated with the following parameters.
In the first one, the renormalization ($\muR$) and factorization ($\muF$) scale were varied by a factor of 0.5, the value of \hdamp was doubled ($2\mt$) and the corresponding Perugia 2012 radiation tune variation was used.  In the second sample,
the renormalization and factorization scales were varied by a factor of 2, while the \hdamp parameter was not changed. The associated variation from the Perugia 2012 tune was used. These choices of parameters have been shown to encompass the cases where $\muR$ and $\muF$ are varied independently, and covered the measured uncertainties of the data for unfolded $t\bar{t}$ distributions~\cite{ATL-PHYS-PUB-2016-004,ATL-PHYS-PUB-2015-002}.
Differences in parton shower and hadronization models were investigated using a sample where \POWHEGBOX was interfaced with \progname{\HERWIGpp} 2.7.1~\cite{Bahr:2008pv} with the UE-EE-5 tune~\cite{ATL-PHYS-PUB-2014-021} and the corresponding CTEQ6L1 PDFs.
The nominal sample was also compared with a sample generated with \progname{\MGMCatNLO}~2.2.1~\cite{Alwall:2014} interfaced with \progname{\HERWIGpp}. A NLO matrix element and CT10 PDF were used for the $t\bar{t}$ hard-scattering process. The parton shower, hadronization and the underlying events were modelled using the \progname{\HERWIGpp}~2.7.1 generator. The UE-EE-5 tune and the corresponding CTEQ6L1 PDF were used.
A functional form was chosen to encompass the differences in the \mll distribution shape between the nominal sample and all these alternative samples. These samples were also used to estimate the 3.5$\%$ extrapolation uncertainty from CRttbar to the SRbTag.
 
Theoretical uncertainties that arise from higher-order contributions to the cross-sections and PDFs and affect the values of the predicted cross-sections for the diboson and single-top backgrounds are considered.
 
\subsubsection*{Signal processes}
Uncertainties related to signal modelling include the uncertainties associated with the initial- and final-state radiation, the modelling of underlying events, the choice of the renormalization and factorization scales, and the parton distribution functions.
None of them has a sizeable effect on the shape of the \mll spectra, but some of them do affect the acceptance in the signal regions.
Uncertainties for the different mass hypotheses were evaluated, but for simplicity, only the largest of the values is used.
In the calculation of the uncertainties, appropriate requirements on the muon and jets kinematics are applied at hadron level for the SRbTag and SRbVeto regions.
 
The factorization and renormalization scales were varied by a factor of two up and down, including correlated and anti-correlated variations, both in \POWHEGBOX and in \progname{MG5\_aMC@NLO}.
For the bb$\X$ process, the largest deviation from the value of the nominal acceptance is considered as the final scale uncertainty (2$\%$ in SRbTag, and 1$\%$ in SRbVeto). A 25$\%$ uncertainty for the acceptance of ggF events in the SRbTag is considered, as well as its anti-correlated effect in the SRbVeto region, following the procedure adopted in Ref.~\cite{ST}.
 
The PDF uncertainties are estimated by taking the envelope of the changes in the acceptance associated with the
PDF4LHC15\_nlo\_100 (PDF4LC15\_nlo\_nf4\_30) ~\cite{Butterworth:2015oua} error set for the ggF (bb\X) signal process. They correspond to changes in acceptances not larger than 1$\%$.
 
Systematic variations of the parameters of the A14 (for bb\X) and AZNLO (for ggF) tunes are used to account
for the uncertainties associated with the initial and final state radiation and the modelling of underlying events.
For bb\X, these uncertainties correspond to changes of 3.8$\%$ and 3.2$\%$ of the acceptance in the SRbTag and SRbVeto regions.
They are associated with a migration of events from one signal region to the other, hence they are treated as anti-correlated between the two signal regions.
For ggF, the uncertainty in the SRbVeto is negligible ($<0.5\%$), while it is 3.8$\%$ in the SRbTag.
 
As discussed in Section~\ref{sec:samples}, the signal samples were simulated with the fast simulation framework of ATLAS, which replaces the full simulation of the electromagnetic and hadronic calorimeters by a parameterized model. No significant differences in acceptance and dimuon invariant mass were observed between a limited number of fast simulation signal samples and samples generated with full Geant4-based simulation, hence no systematic uncertainty is considered for the use of fast simulation.
 
The current result is dominated by the statistical uncertainty of the data (which accounts for 66$\%$ of the total uncertainty on the fitted value of the signal cross-section), and the systematic uncertainty due to the finite size of the simulated background samples (which accounts for 25$\%$ of the total uncertainty).
Amongst the remaining systematic uncertainties, the modelling of the shape of the \mll distribution for the $Z$+jets and top--antitop backgrounds dominates (3$\%$ of the total uncertainty) followed by the muon identification efficiency (1.5$\%$). All other considered sources of experimental and theoretical uncertainties (which individually have an impact $<0.4\%$) have a combined contribution of less than 5$\%$ to the total uncertainty on the upper limit.
% End of text imported from the .//ANA-HIGG-2017-10-PAPER-systematics.tex input file
% The next lines are included from the .//ANA-HIGG-2017-10-PAPER-results.tex input file
\section{Results}
\label{sec:result}
The observed $p$-values as a function of \mx, obtained applying the statistical analysis described in Section~\ref{sec:stat}, are shown in Figure~\ref{Cs_p0_SRbTag} for the \bbx-only fit and in Figure~\ref{Cs_p0_SRbVeto} for the ggF-only fit.
The lines in each figure correspond to separate fits. The dotted line represents the $p$-value for a fit including only SRbTag and all control regions (CRs), the dashed line represents the $p$-value for a fit including only SRbVeto+CRs, while the solid line corresponds to the combined SRbTag+SRbVeto+CRs fit.
The dotted lines in the two figures show similar behaviour and so do the dashed lines,
while the combined $p$-values are calculated for separate signal production modes implying very different contributions from the SRbVeto and SRbTag regions, which lead to substantially different limits.
The largest excess of events above the expected background is observed for the \bquark
associated production at about $\mx=480\,\GeV$ and amounts to a
local significance of $2.3\sigma$, which becomes $0.6\sigma$
after considering the look-elsewhere effect over the mass range
0.2--1.0~TeV. The look-elsewhere effect is estimated using the method described in
Ref.~\cite{Gross:2010qma}.
\begin{figure}[!ht]
\centering
\subfloat[]{\includegraphics[width=0.45\linewidth]{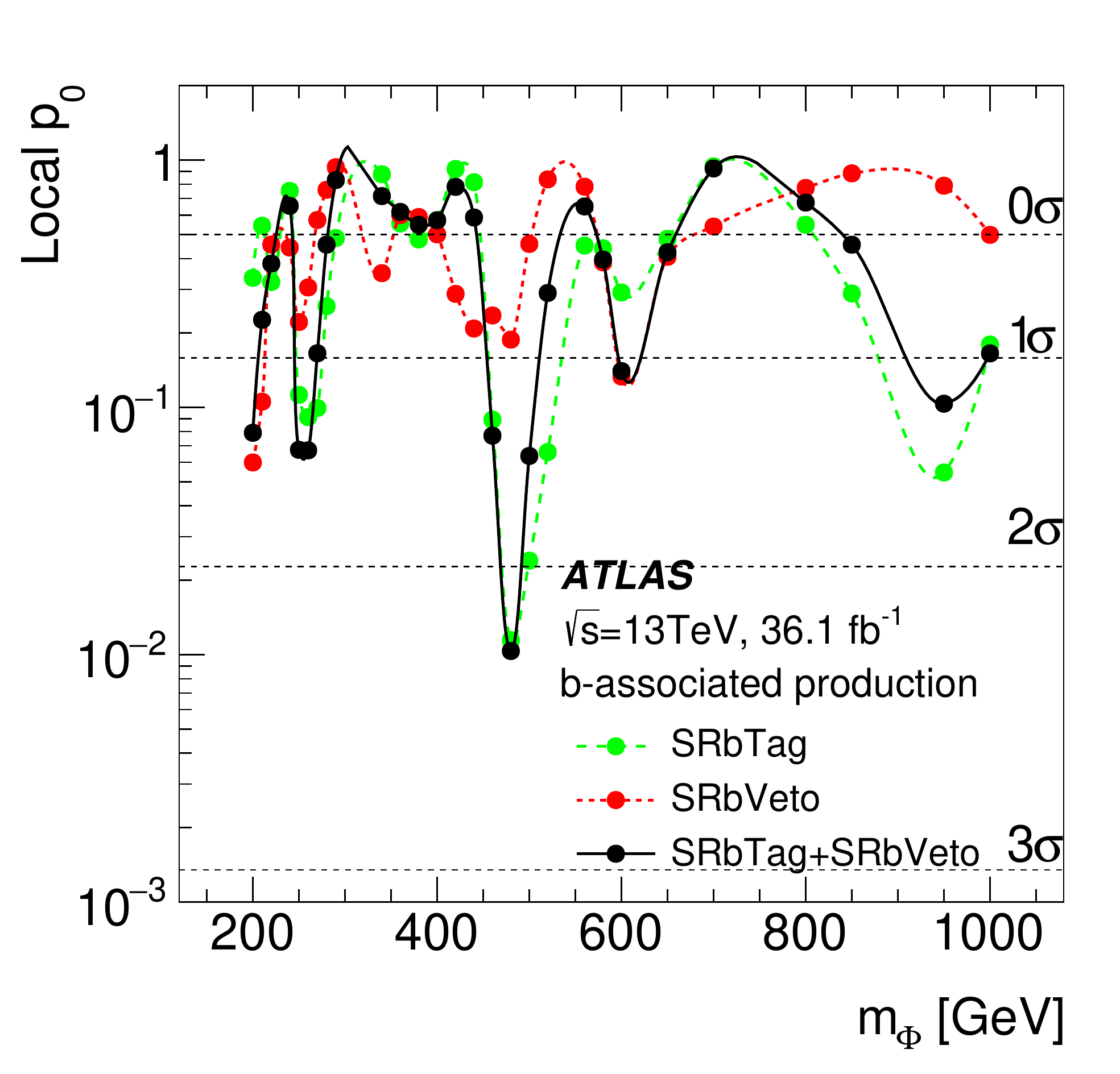}\label{Cs_p0_SRbTag} }
\subfloat[]{\includegraphics[width=0.45\linewidth]{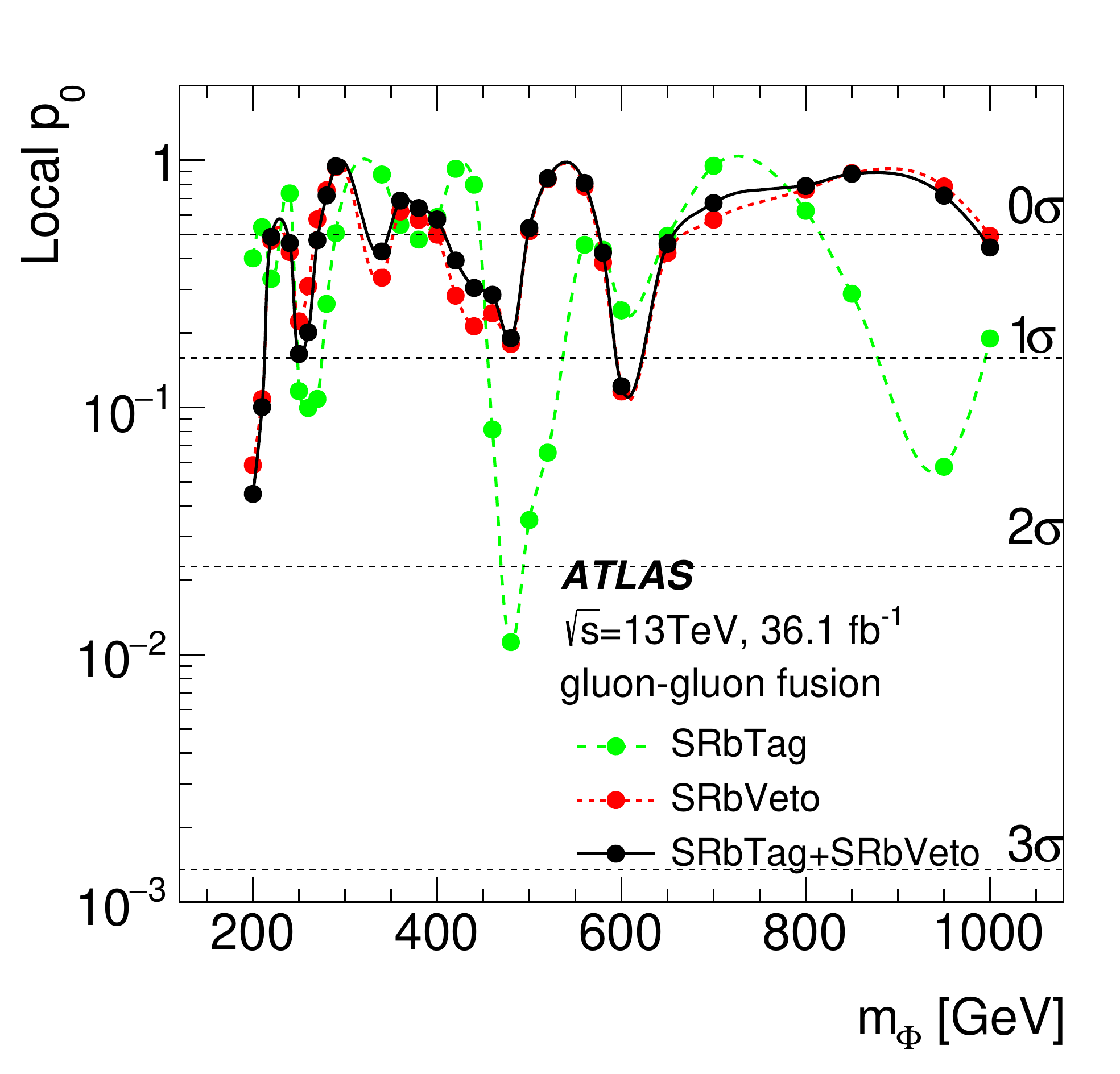}  \label{Cs_p0_SRbVeto}}
\caption{
Observed $p$-values as a function of \mx for (a) a \bbx-only fit  and (b) a ggF-only fit. The dotted line corresponds to the data in the SRbTag only, the dashed line to the data in SRbVeto only, and the solid curve to the combination of the two data categories. All three fits include data from all control regions.
}
\label{fig:Cs_p0}
\end{figure}
 
Since the data are in agreement with the predicted backgrounds 
the results are given in terms of exclusion limits. These are set using the modified frequentist CL$_{\textrm s}$ method~\cite{Read:2002hq}. Upper limits
on the cross section times branching ratio for a massive scalar particle are set at the 95\% confidence level (CL)
as a function of the particle mass.
The upper limits at 95\% CL on $\sigma_{\X}\times\mathcal{B}(\X\to\mu\mu)$ (where $\sigma_{\X}$ is the cross section and  $\mathcal{B}$ is the branching ratio) assume the natural width of the resonance to be
negligible compared to the experimental resolution, and they cover the mass range
0.2--1.0~TeV. In Figure~\ref{Cs_upperlimits_fbbH100} and~\ref{Cs_upperlimits_fbbH0} the 95\% CL upper limits are
shown for  \bquark associated production and gluon--gluon fusion, respectively. The expected limits are in the ranges 1.3--\SI{25}{\fb} and 1.8--\SI{25}{\fb}, respectively.
The observed limits are in the ranges 1.9--\SI{41}{\fb} and 1.6--\SI{44}{\fb}, respectively.
Both the expected and observed limits are calculated in the asymptotic approximation~\cite{Cowan:2010js}, which was verified with MC pseudo-experiments for the $\mx=1000$~GeV hypothesis.
The upper limit on the ggF production mode is dominated by the SRbVeto, while for the \bbx production mode both the SRbTag and SRbVeto regions contribute to the result because of the relative acceptance of the \bbx production mode in the two regions.
Limits shown in Figure~\ref{fig:Cs_upperlimits} are obtained in simultaneous fits of the SRbVeto and SRbTag data regions and correspond to the different signal production modes.
Usage of the same data events in both the \bbx-only and the ggF-only fits explains correlations in the behaviour of the observed limits.
 
\begin{figure}[!ht]
\centering
\subfloat[]{\includegraphics[width=0.4\linewidth]{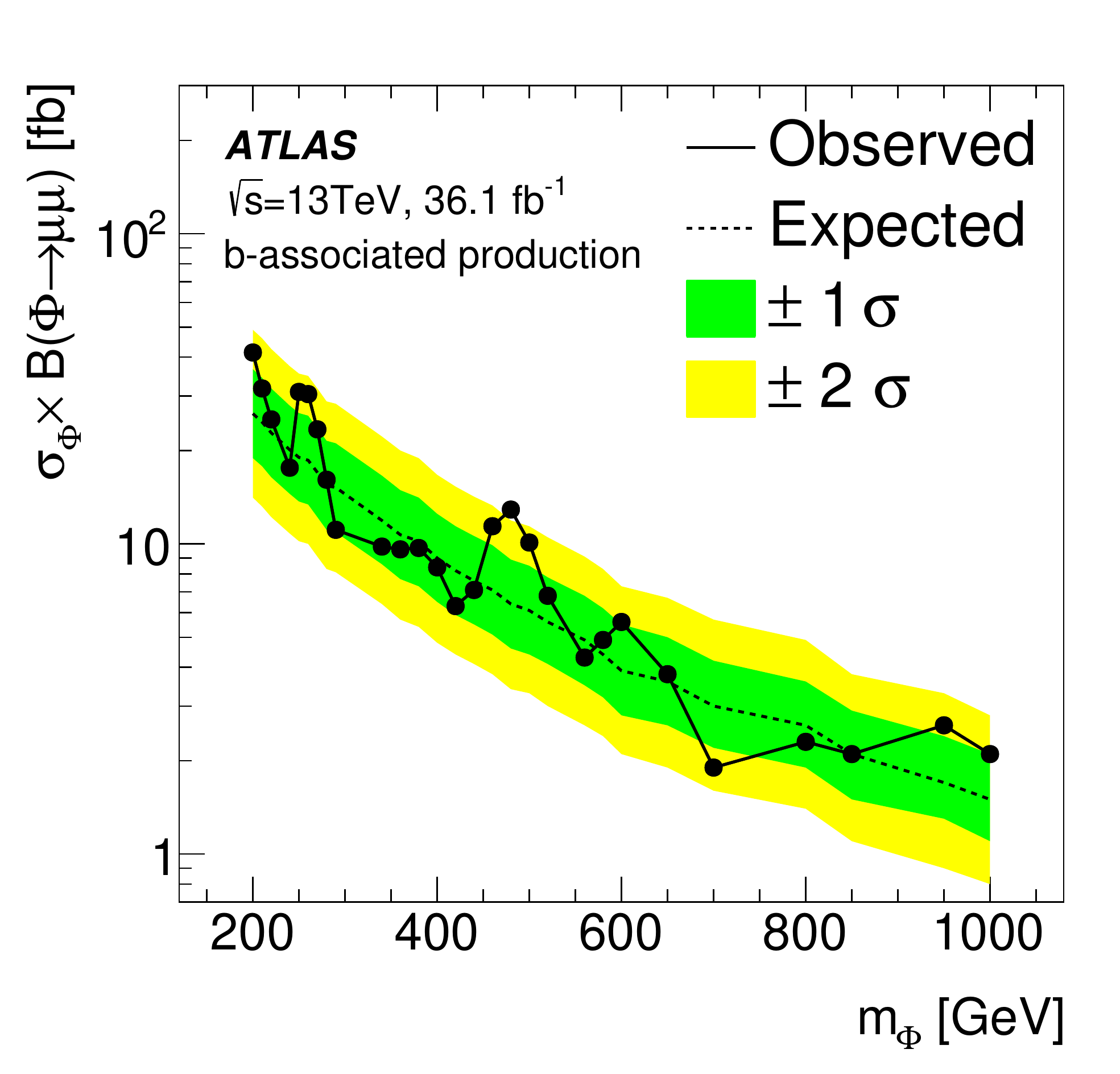}\label{Cs_upperlimits_fbbH100}}
\subfloat[]{\includegraphics[width=0.4\linewidth]{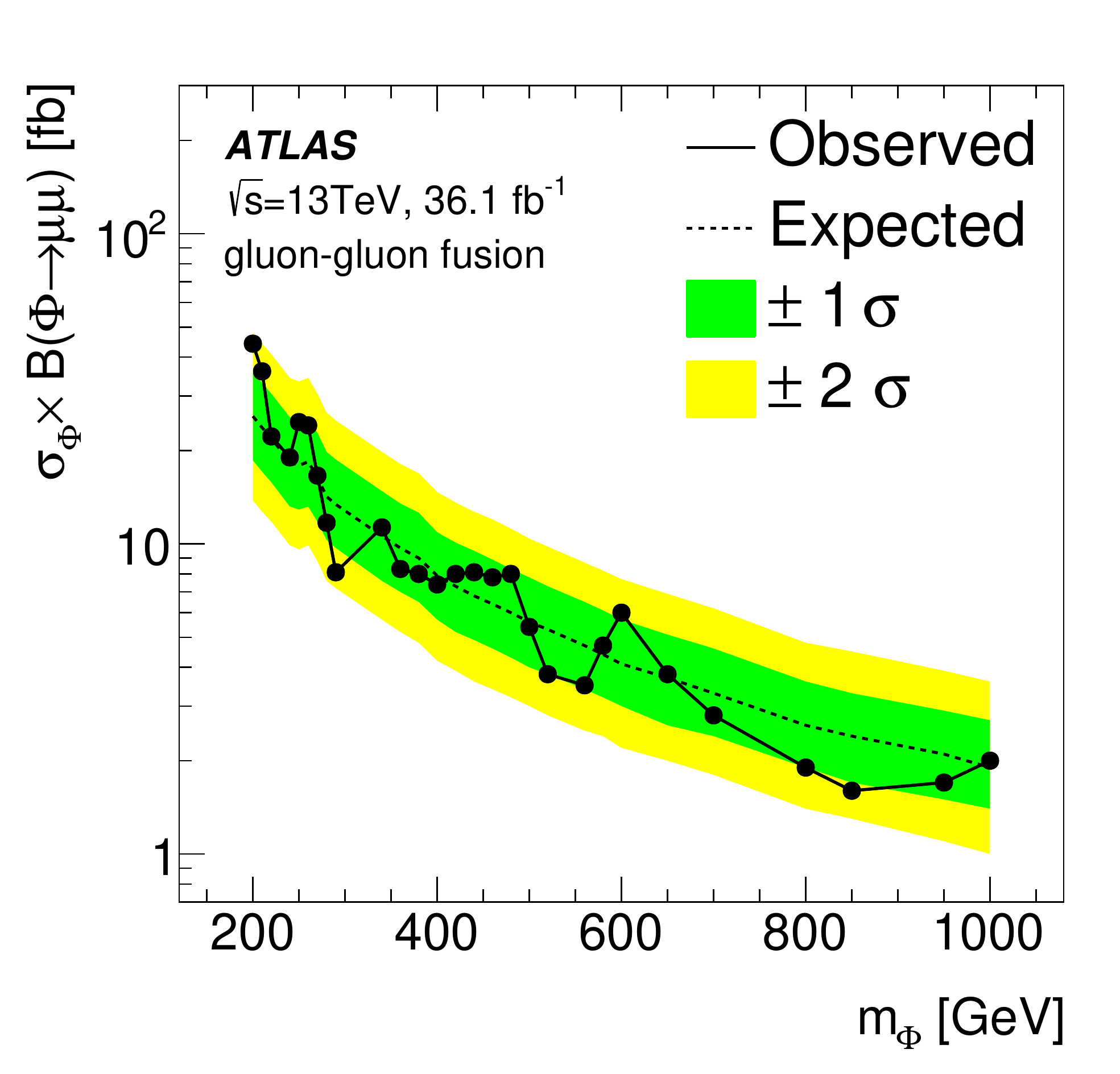}  \label{Cs_upperlimits_fbbH0}  }
\caption{The observed and expected 95\% CL upper limits on the production cross section times branching ratio
for a massive scalar resonance produced via (a) \bquark associated production and (b) gluon–-gluon fusion.}
\label{fig:Cs_upperlimits}
\end{figure}
 
Figures~\ref{Cs_mssm_obs} and~\ref{Cs_mssm_exp} show the observed
and expected 95\% CL upper limits on the production cross section
times branching ratio for $\X\to\mu\mu$ as a function of the
fractional contribution from \bquark associated production
($\sigma_{\text{\bbx}}/[\sigma_\text{{\bbx}} + \sigma_{\text{ggF}}]$) and the scalar resonance mass. These fractions are derived assuming
that other production mechanisms do not affect the result.
 
\begin{figure}[!ht]
\centering
\subfloat[]{\includegraphics[width=0.4\linewidth]{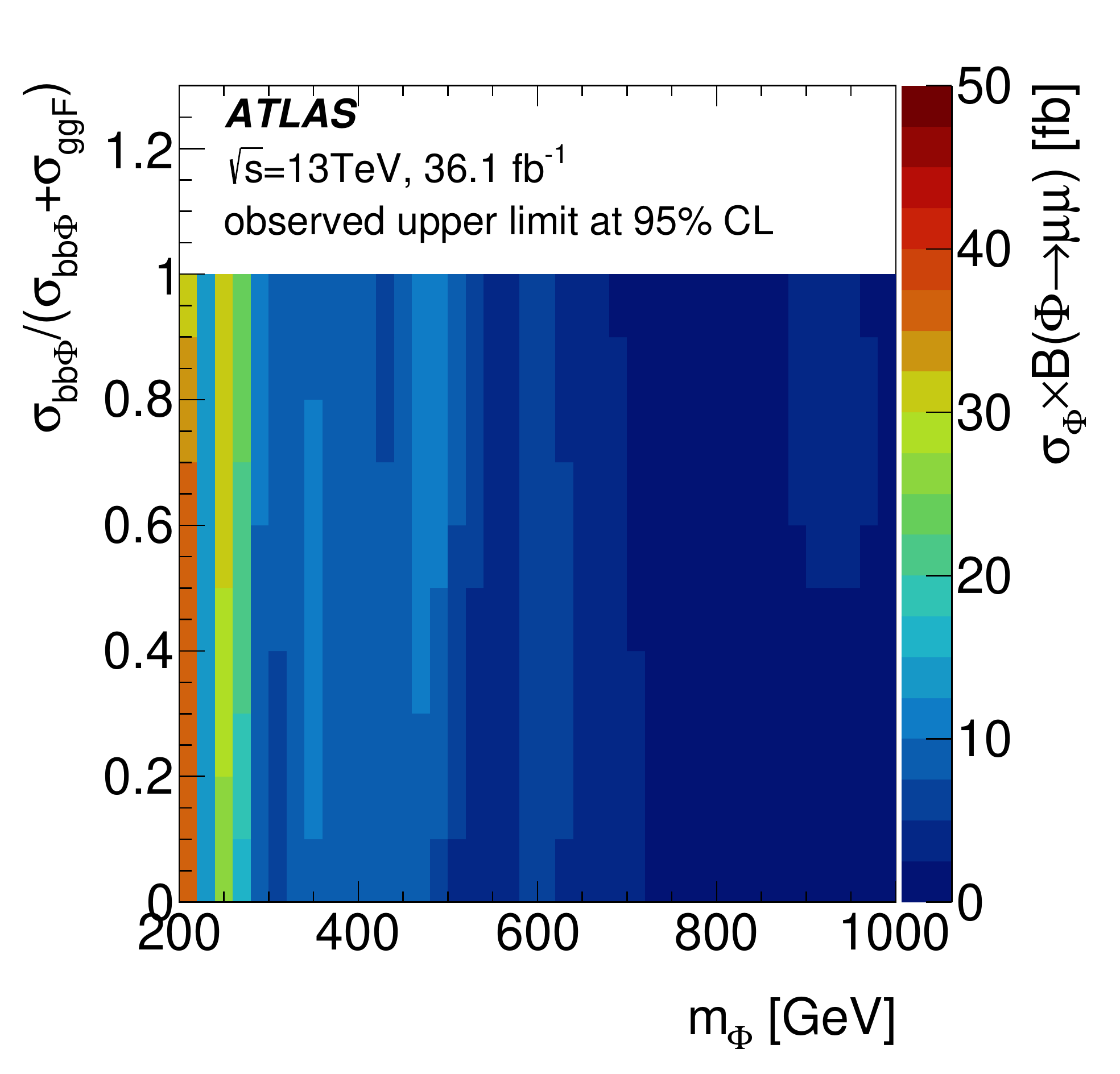}\label{Cs_mssm_obs}}
\subfloat[]{\includegraphics[width=0.4\linewidth]{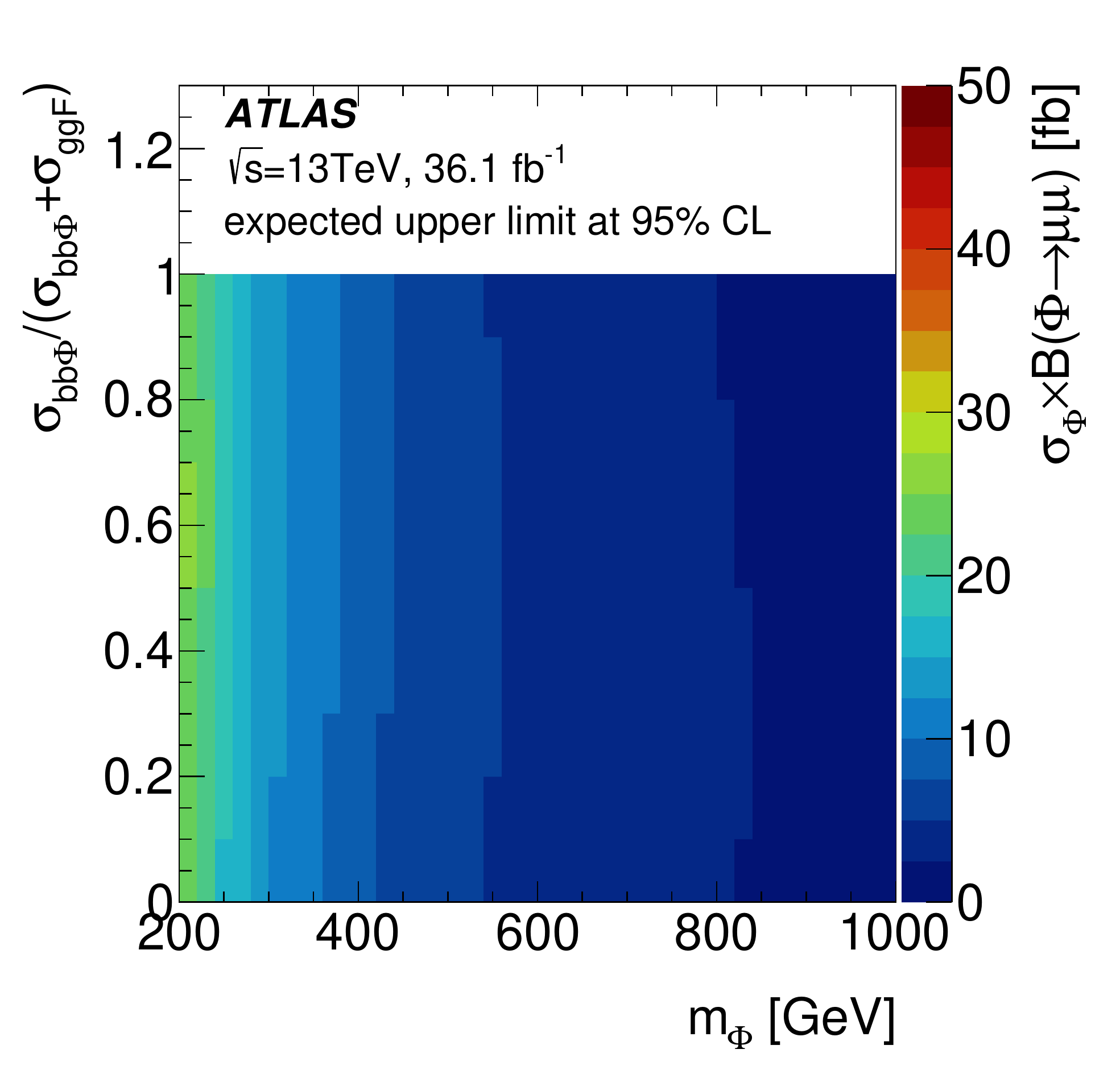}\label{Cs_mssm_exp}}
\caption{The (a) observed and (b) expected 95\% CL upper limit on the production cross section times branching ratio for $\X\to\mu\mu$ as a function of the fractional contribution from \bquark associated production and the scalar boson mass.
}
\label{fig:Cs_mssm}
\end{figure}
 
A complete set of tables and figures are available at the Durham HepData repository~\cite{hepdata}.

\FloatBarrier
% End of text imported from the .//ANA-HIGG-2017-10-PAPER-results.tex input file
% The next lines are included from the .//ANA-HIGG-2017-10-PAPER-conclusion.tex input file
\section{Conclusion}
\label{sec:conclusion}
 
The ATLAS detector at the LHC has been used to search for a massive scalar resonance decaying into two opposite-sign muons,
produced via \bquark associated production or via gluon--gluon fusion, assuming that the natural width of the resonance is  negligible compared to the experimental resolution.
The search is conducted with 36.1~$\ifb$ of \pp collision data at $\rts=13$ $\TeV$, recorded during 2015 and 2016.
The observed dimuon invariant mass spectrum is consistent with the Standard Model prediction
within uncertainties for events both with and without a \btagged jet over the 0.2--1.0~$\TeV$ range.
The observed upper limits at 95\% confidence level on the cross-section times branching
ratio for \bquark associated production and gluon--gluon fusion are between 1.9 and 41 fb and 1.6 and 44 fb respectively,
which is consistent with expectations.

% End of text imported from the .//ANA-HIGG-2017-10-PAPER-conclusion.tex input file
 
\section*{Acknowledgements}
 
% The next lines are included from the .//acknowledgements/Acknowledgements.tex input file
 
We thank CERN for the very successful operation of the LHC, as well as the
support staff from our institutions without whom ATLAS could not be
operated efficiently.
 
We acknowledge the support of ANPCyT, Argentina; YerPhI, Armenia; ARC, Australia; BMWFW and FWF, Austria; ANAS, Azerbaijan; SSTC, Belarus; CNPq and FAPESP, Brazil; NSERC, NRC and CFI, Canada; CERN; CONICYT, Chile; CAS, MOST and NSFC, China; COLCIENCIAS, Colombia; MSMT CR, MPO CR and VSC CR, Czech Republic; DNRF and DNSRC, Denmark; IN2P3-CNRS, CEA-DRF/IRFU, France; SRNSFG, Georgia; BMBF, HGF, and MPG, Germany; GSRT, Greece; RGC, Hong Kong SAR, China; ISF and Benoziyo Center, Israel; INFN, Italy; MEXT and JSPS, Japan; CNRST, Morocco; NWO, Netherlands; RCN, Norway; MNiSW and NCN, Poland; FCT, Portugal; MNE/IFA, Romania; MES of Russia and NRC KI, Russian Federation; JINR; MESTD, Serbia; MSSR, Slovakia; ARRS and MIZ\v{S}, Slovenia; DST/NRF, South Africa; MINECO, Spain; SRC and Wallenberg Foundation, Sweden; SERI, SNSF and Cantons of Bern and Geneva, Switzerland; MOST, Taiwan; TAEK, Turkey; STFC, United Kingdom; DOE and NSF, United States of America. In addition, individual groups and members have received support from BCKDF, CANARIE, CRC and Compute Canada, Canada; COST, ERC, ERDF, Horizon 2020, and Marie Sk{\l}odowska-Curie Actions, European Union; Investissements d' Avenir Labex and Idex, ANR, France; DFG and AvH Foundation, Germany; Herakleitos, Thales and Aristeia programmes co-financed by EU-ESF and the Greek NSRF, Greece; BSF-NSF and GIF, Israel; CERCA Programme Generalitat de Catalunya, Spain; The Royal Society and Leverhulme Trust, United Kingdom.
 
The crucial computing support from all WLCG partners is acknowledged gratefully, in particular from CERN, the ATLAS Tier-1 facilities at TRIUMF (Canada), NDGF (Denmark, Norway, Sweden), CC-IN2P3 (France), KIT/GridKA (Germany), INFN-CNAF (Italy), NL-T1 (Netherlands), PIC (Spain), ASGC (Taiwan), RAL (UK) and BNL (USA), the Tier-2 facilities worldwide and large non-WLCG resource providers. Major contributors of computing resources are listed in Ref.~\cite{ATL-GEN-PUB-2016-002}.
 
% End of text imported from the .//acknowledgements/Acknowledgements.tex input file

\printbibliography

\clearpage \input{atlas_authlist}

\end{document}

%% file: atlas_authlist.tex
% ATLAS Collaboration author list
% Reference date of HIGG-2017-10 is 2018-08-21
% Author list last updated on date 13-NOV-18
% Data extracted on 13-Nov-2018 for paper reference HIGG-2017-10
% at 7:12am
 
\begin{flushleft}
{\Large The ATLAS Collaboration}

\bigskip

M.~Aaboud$^\textrm{\scriptsize 34d}$,    
G.~Aad$^\textrm{\scriptsize 99}$,    
B.~Abbott$^\textrm{\scriptsize 125}$,    
D.C.~Abbott$^\textrm{\scriptsize 100}$,    
O.~Abdinov$^\textrm{\scriptsize 13,*}$,    
B.~Abeloos$^\textrm{\scriptsize 129}$,    
D.K.~Abhayasinghe$^\textrm{\scriptsize 91}$,    
S.H.~Abidi$^\textrm{\scriptsize 164}$,    
O.S.~AbouZeid$^\textrm{\scriptsize 39}$,    
N.L.~Abraham$^\textrm{\scriptsize 153}$
% ...
\bigskip
\\

% ...
$^{13}$Institute of Physics, Azerbaijan Academy of Sciences, Baku; Azerbaijan.\\
% ...
$^{34}$$^{(a)}$Facult\'e des Sciences Ain Chock, R\'eseau Universitaire de Physique des Hautes Energies - Universit\'e Hassan II, Casablanca;$^{(b)}$Centre National de l'Energie des Sciences Techniques Nucleaires (CNESTEN), Rabat;$^{(c)}$Facult\'e des Sciences Semlalia, Universit\'e Cadi Ayyad, LPHEA-Marrakech;$^{(d)}$Facult\'e des Sciences, Universit\'e Mohamed Premier and LPTPM, Oujda;$^{(e)}$Facult\'e des sciences, Universit\'e Mohammed V, Rabat; Morocco.\\
% ...
$^{39}$Niels Bohr Institute, University of Copenhagen, Copenhagen; Denmark.\\
% ...
$^{91}$Department of Physics, Royal Holloway University of London, Egham; United Kingdom.\\
% ...
$^{99}$CPPM, Aix-Marseille Universit\'e, CNRS/IN2P3, Marseille; France.\\
$^{100}$Department of Physics, University of Massachusetts, Amherst MA; United States of America.\\
% ...
$^{125}$Homer L. Dodge Department of Physics and Astronomy, University of Oklahoma, Norman OK; United States of America.\\
% ...
$^{129}$LAL, Universit\'e Paris-Sud, CNRS/IN2P3, Universit\'e Paris-Saclay, Orsay; France.\\
% ...
$^{153}$Department of Physics and Astronomy, University of Sussex, Brighton; United Kingdom.\\
% ...
$^{164}$Department of Physics, University of Toronto, Toronto ON; Canada.\\
% ...
$^{a}$ Also at Borough of Manhattan Community College, City University of New York, NY; United States of America.\\
$^{b}$ Also at California State University, East Bay; United States of America.\\
$^{c}$ Also at Centre for High Performance Computing, CSIR Campus, Rosebank, Cape Town; South Africa.\\
$^{d}$ Also at CERN, Geneva; Switzerland.\\
$^{e}$ Also at CPPM, Aix-Marseille Universit\'e, CNRS/IN2P3, Marseille; France.\\
% ...
$^{*}$ Deceased

\end{flushleft}

% Created with Glance <Atlas.Glance@cern.ch>